\documentclass[12pt]{article}
\usepackage{graphicx}

\newcommand{\Frac}[2]{\frac{\displaystyle #1}{\displaystyle #2}}
\newcommand{\beq}{\begin{equation}}
\newcommand{\eeq}{\end{equation}}
\newcommand{\beqn}{\begin{eqnarray}}
\newcommand{\eeqn}{\end{eqnarray}}
\newcommand{\beqns}{\begin{eqnarray*}}
\newcommand{\eeqns}{\end{eqnarray*}}

\begin{document}
\begin{titlepage}
\begin{center}

\hfill USTC-ICTS-02-04       \\
\hfill  December 2002

\vspace{2.5cm}

{\Large {\bf  Charge asymmetry in $K^\pm\to\pi^\pm \gamma \gamma$
induced by the electromagnetic penguin operators
\\}} \vspace*{1.0cm}
 {  Dao-Neng Gao$^\dagger$} \vspace*{0.3cm} \\
{\it\small Department of Astronomy and Applied Physics, University
of Science and Technology of China\\ Hefei, Anhui 230026 China
\\and\\ Interdisciplinary Center for Theoretical Study, University
of Science and Technology of China\\ Hefei, Anhui 230026 China }

\vspace*{1cm}
\end{center}
\begin{abstract}
\noindent The CP-violating charge asymmetry in decays
$K^\pm\to\pi^\pm\gamma\gamma$, which is induced by the
electromagnetic penguin operators, has been studied both in the
standard model and in its extensions. Because of a large
enhancement of the Wilson coefficients of the electromagnetic
penguin operators in the supersymmetric extensions of the standard
model, a significant upper bound on this charge asymmetry could be
expected, and thus high-precision measurements of this interesting
CP-violating quantity might explore new physics effects beyond the
standard model.
\end{abstract}

\vfill
\noindent

$^{\dagger}$ E-mail:~gaodn@ustc.edu.cn
\end{titlepage}

Rare kaon decays provide a very useful laboratory both to test the
standard model (SM) and to explore new physics beyond it
\cite{BK00, BB01, DI98, DE86}. It is of particular interest to
study the CP-violating effects, which arises in such weak decays
from dimension-five operators, including the electromagnetic and
chromomagnetic penguin operators (EMO and CMO) since the CP
violation induced by these operators is suppressed in the SM;
however, it could be enhanced in its extensions \cite{BCIRS99,
CIP99, DIM99, HMPV99,TV00, BJKP01, Messina02, DG02}. On the other
hand, present experiments, HyperCP \cite{HyperCP}, and future NA48
experiments \cite{NA48b}, are going to substantially improve the
present limits on the Wilson coefficients of these operators by
studying CP-violating charge asymmetries in charged kaon decays,
such as $K^\pm \rightarrow (3 \pi)^\pm$, $K^\pm \rightarrow
\pi^\pm \ell^+ {\ell}^-$, as well as one-photon or two-photon
radiative decays. It is expected that charged kaon decays could be
an ideal framework to explore direct CP violation, or CP violation
of pure $\Delta S=1$ origin \cite{DI98, CIP99, DIM99, Messina02,
DG02, Retico02}. The purpose of this paper is devoted to the
analysis of the CP-violating charge asymmetry induced by the EMO
in the two-photon radiative charged kaon decay, $K^\pm\to \pi^\pm
\gamma \gamma$, both in the SM and in its possible extensions.

The general weak effective Hamiltonian, contributed by the EMO and
CMO, can be written as \cite{BCIRS99} \beq\label{HMO} {\cal
H}_{\rm eff}=C_\gamma^+(\mu) Q_\gamma^+(\mu)+ C_\gamma^-(\mu)
Q_\gamma^-(\mu)+C_g^+(\mu) Q_g^+(\mu) +C_g^-(\mu) Q_g^-(\mu)+ {\rm
H.c.},\label{cpg} \eeq where $C^\pm_{\gamma,~g}$ are the Wilson
coefficients and \beqn Q_\gamma^\pm=\Frac{e
Q_d}{16\pi^2}(\bar{s}_L\sigma_{\mu\nu}d_R\pm
\bar{s}_R\sigma_{\mu\nu}d_L)F^{\mu\nu},\label{EMO}\\
Q_g^\pm=\Frac{g_s}{16\pi^2}(\bar{s}_L\sigma_{\mu\nu}t_a d_R\pm
\bar{s}_R\sigma_{\mu\nu}t_a d_L) G_a^{\mu\nu}.\label{CMO} \eeqn
Here $Q_d=-1/3$, and $\sigma_{\mu\nu}=i/2[\gamma_\mu,
\gamma_\nu]$. Note that eq. (\ref{HMO}) with complex Wilson
coefficients $C^\pm_{\gamma,~g}$ could lead to new flavor
structures beyond the SM, which generally depart from minimal
flavor violation \cite{GGIS02}. It is easy to see that the SM
structure, $SU(2)_L \times U(1)_Y$, will impose the following
chiral suppression for these operators \cite{RPS93, BME94}:
 \beqn {\cal
H}^{\rm SM}_{\rm eff}=
\Frac{G_F}{\sqrt{2}}V_{td}V^*_{ts}\left[C_{11}\frac{g_s}{8\pi^2}
(m_d\bar{s}_L\sigma_{\mu\nu}t_a
d_R+m_s\bar{s}_R\sigma_{\mu\nu}t_a d_L)G_a^{\mu\nu}\nonumber \right.\\
\left.+C_{12}\frac{e}{8\pi^2}(m_d\bar{s}_L\sigma_{\mu\nu}d_R+m_s\bar{s}_R
\sigma_{\mu\nu}d_L)F^{\mu\nu}\right]+ {\rm H.c.}, \label{SMHMO}
\eeqn with \beqn C_{11}(m_W)=\Frac{3x^2}{2(1-x)^4}{\rm
ln}~x-\Frac{x^3-5x^2-2x}{4(1-x)^3},\label{SMC11}\\
C_{12}(m_W)=\Frac{x^2(2-3x)}{2(1-x)^4}{\rm
ln}~x-\Frac{8x^3+5x^2-7x}{12(1-x)^3}, \label{SMC12} \eeqn where
$x=m_t^2/m_W^2$ and $t_a$ are the $SU(3)$-matrices. However, as we
shall see, new flavor structures, for instance, from the
supersymmetric extensions of the SM,  allow us to avoid the chiral
suppression for the operators in eq. (\ref{SMHMO}) \cite{GGMS96}.

It has been known that the $K^\pm\to\pi^\pm\gamma\gamma$
transitions are dominated by long distance effects \cite{EPR88,
E787}. Within the framework of chiral perturbation theory
\cite{EPR88, KMW90}, $K^\pm\to\pi^\pm \gamma\gamma$ receives the
first non-vanishing contribution at $O(p^4)$ including both loops
and anomalous and non-anomalous couterterms. The general $O(p^4)$
amplitude of the decay can be decomposed in the following way
\cite{EPR88} \beqn\label{amp0}
&&M(K^+(k)\to\pi^+(p)\gamma(q_1,\epsilon_1)\gamma(q_2,\epsilon_2))\nonumber\\
&&\;\;\;\;\;\;=\epsilon_{1\mu}(q_1)\epsilon_{2\nu}(q_2)\left[\frac{A(y,z)}{m_K^2}(q_2^\mu
q_1^\nu-q_1\cdot q_2
g^{\mu\nu})+\frac{C(y,z)}{m_K^2}\varepsilon^{\mu\nu\alpha\beta}q_{1\alpha}
q_{2\beta}\right],\eeqn with \beq y={k\cdot(q_1-q_2)\over
m_K^2},\;\;\;\;\;\;\;\;z={(q_1+q_2)^2\over m_K^2}.\eeq The
physical region in the dimensionless variables $y$ and $z$ is
given by \beq 0\leq|y|\leq
\frac{1}{2}\lambda^{1/2}(1,z,r_\pi^2),\;\;\;\; 0\leq
z\leq(1-r_\pi)^2,\eeq where $r_\pi=m_\pi/m_K$ and
$\lambda(a,b,c)=a^2+b^2+c^2-2(ab+ac+bc).$ Note that the invariant
amplitudes $A(y,z)$ from loops and non-anomalous counterterms, and
$C(y,z)$ from anomalous counterterms have to be symmetric under
the interchange of $q_1$ and $q_2$ as required by Bose symmetry.
In the SM, the $O(p^4)$ amplitude for $A(y,z)$ has been given in
Ref. \cite{EPR88}, which is \beq\label{amp1} A(y,z)=\frac{G_8m_K^2
\alpha_{\rm EM}}{2\pi z}\left[(r_\pi^2-1-z)F\left({z\over r_\pi^2
}\right)+(1-r_\pi^2-z)F(z)+\hat{c} z\right],
 \eeq where $\alpha_{\rm EM}=e^2/4\pi$, and $|G_8|=9.2\times 10^{-6}$ GeV$^{-2}$.
We do not display the explicit expression for $C(y,z)$ since it is
irrelevant to the present discussion. $F(z/r_\pi^2)$ and $F(z)$
are generated from $\pi$ and $K$ loop diagrams respectively, which
could be defined as \beq\label{loop}F(x)=\left\{\begin{array}{cc}
1-\Frac{4}{x}\arcsin^2\left(\Frac{\sqrt{x}}{2}\right) & x\leq
4,\\\\1+\Frac{1}{x}\left(\ln{\Frac{1-\sqrt{1-4/x}}{1+\sqrt{1-4/x}}}+i\pi\right)^2&x\geq
4.\end{array}\right. \eeq $\hat{c}$ in eq. (\ref{amp1}) is from
$O(p^4)$ non-anomalous local counterterms \beq
\hat{c}=\frac{128\pi^2}{3}\left[3 (L_9+L_{10})+N_{14}-N_{15}-2
N_{18} \right],\eeq where $L_9$ and $L_{10}$ are couplings in the
$O(p^4)$ strong chiral lagrangian \cite{GL84} and $N_{i}$ ($i$=14,
15,18) are couplings in the $O(p^4)$ weak chiral lagrangian
\cite{KMW90}.

From eq. (\ref{loop}) it is obvious that the $\pi$ loop
contribution, which is proportional to $F(z/r_\pi^2)$ in eq.
(\ref{amp1}), will generate a CP invariant absorptive part. Thus
if $\hat{c}$ has a non-vanishing phase, the interference between
these two parts will lead to the charge asymmetry in
$K^\pm\to\pi^\pm\gamma\gamma$ as follows \cite{EPR88,DEIN95}
\beqn\label{asy0}\Frac{\delta\Gamma}{2\Gamma}&=&
\Frac{|\Gamma(K^+\to\pi^+\gamma\gamma)-\Gamma(K^-\to\pi^-\gamma\gamma)|}
{\Gamma(K^+\to\pi^+\gamma\gamma)+\Gamma(K^-\to\pi^-\gamma\gamma)},
\\\nonumber\\
\label{asy1}\delta\Gamma
&=&|\Gamma(K^+\to\pi^+\gamma\gamma)-\Gamma(K^-\to\pi^-\gamma\gamma)|\nonumber\\
&=&\Frac{{\rm Im}\hat{c}|G_8\alpha_{\rm EM}|^2
m_{K}^5}{2^{10}\pi^5}\int^{(1-r_\pi)^2}_{4r_\pi^2}dz~\lambda^{1/2}(1,z,r_\pi^2)(r_\pi^2-1-z)z
~{\rm Im}F(z/r_\pi^2). \eeqn Thus information about the imaginary
part of $\hat{c}$ is relevant to the estimate of $\delta \Gamma$.
It is clear that $L_9$ and $L_{10}$ cannot contribute to it. In
the SM, ${\rm Im}\hat{c}$ is therefore governed by the weak
coupling combination $N_{14}-N_{15}-2 N_{18}$ due to the
Cabbibo-Kobayashi-Maskawa (CKM) phase \cite{CKM}. This point has
been explored in the past literature, and some small values of the
charge asymmetry in $K^\pm\to\pi^\pm\gamma\gamma$ have been
obtained in Refs. \cite{DEIN95} and \cite{DI98, CDM93},
respectively; \beqn\label{updated}
\left(\frac{\delta\Gamma}{2\Gamma}\right)^{\rm SM}\ll10^{-3}~
\cite{DEIN95},\eeqn and \beqn
\left(\frac{\delta\Gamma}{2\Gamma}\right)^{\rm SM
}<10^{-4}~\cite{DI98}.\label{unitarity}\eeqn We would like to give
some remarks here. First, eq. (\ref{updated}) is the updated
version of the result given in Ref. \cite{EPR88}, and a vanishing
imaginary part of $N_{14}-N_{15}-2 N_{18}$ has been predicted by
the authors of Ref. \cite{BP93} using $1/N_C$ analysis. Second, in
deriving eq. (\ref{unitarity}), the $O(p^6)$ unitarity corrections
\cite{CDM93, DP96} from the physical $K^\pm\to 3\pi$ vertex to
$K^\pm\to\pi^\pm \gamma\gamma$ have been taken into account. As an
order-of-magnitude estimate for the charge asymmetry in
$K^\pm\to\pi^\pm\gamma\gamma$ induced by the EMO, in the present
paper we do not consider the $O(p^6)$ unitarity contributions
since, as pointed out in Ref. \cite{CDM93}, it does not alter
significantly the value of the charge asymmetry obtained at
$O(p^4)$ in chiral perturbation theory.

Let us now try to delve into the analysis of the CP-violating
charge asymmetry in $K^\pm\to\pi^\pm\gamma \gamma$ arising from
the EMO formulated in the general effective Hamiltonian of eq.
(\ref{HMO}). It will be shown below that, this asymmetry is very
small in the SM, however, it could be significant in the
supersymmetric scenarios beyond the SM by comparison with those
given in eqs. (\ref{updated}) and (\ref{unitarity}). Using the
same way given in Ref. \cite{TV00}, we first construct the
effective lagrangian that represents the EMO. In fact there are
many possible chiral realization of these operators. To the
purpose of this paper, the leading order $O(p^4)$ realizations of
the EMO, which are relevant to $K\to \pi \gamma \gamma$
transitions, could be expressed as \beq\label{effL0} {\cal L}=a_1
{\cal L}_1+a_2 {\cal L}_2,\eeq \beq\label{effL1} {\cal
L}_1=\frac{e Q_d}{16\pi^2}C_\gamma^\pm\langle \lambda
U(F_{L\mu\nu}+U^\dagger F_{R\mu\nu}U)\pm\lambda
(F_{L\mu\nu}+U^\dagger F_{R\mu\nu}U)U^\dagger\rangle
F^{\mu\nu}+{\rm H.c.}, \eeq and \beq\label{effL2} {\cal L}_2=i
\frac{e Q_d}{16\pi^2}C_\gamma^\pm\langle \lambda U L_\mu L_\nu\pm
\lambda L_\mu L_\nu U^\dagger\rangle F^{\mu\nu}+{\rm H.c.,}\eeq
where $a_1$ and $a_2$ are unknown coupling constants, $L_\mu=i
U^\dagger D_\mu U$, and $(\lambda)_{ij}=\delta_{3i}\delta_{2i}$.
We use the standard notation in chiral perturbation theory
\cite{Pich98}, $F_{L\mu\nu}=F_{R\mu\nu}=eQF_{\mu\nu}$, $D_\mu
U=\partial_\mu U-ie [Q,U]A_\mu$, and $Q={\rm diag}(2,-1,-1)/3$.
$U$ is a unitary $3\times 3$ matrix with det~$U$=1, which collects
the Goldstone meson fields ($\pi$, $K$, and $\eta$) as follows
\beqn\label{Phi}
&&U={\rm exp}(i\sqrt{2}\Phi/f_\pi),\nonumber\\\nonumber\\
&&\Phi=\frac{1}{\sqrt{2}}\lambda^a\phi_a(x)=\left(\begin{array}{ccc}
\frac{\pi^0}{\sqrt{2}}+\frac{\eta_8}{\sqrt{6}} &\; \pi^+ &\;K^+ \\ \; \\
\pi^- &\;-\frac{\pi^0}{\sqrt{2}}+\frac{\eta_8}{\sqrt{6}} &\; K^0\\ \; \\
K^- &\;\bar{K}^0 & \; -\frac{2\eta_8}{\sqrt{6}}\\
\end{array}\right) \ , \eeqn
where the $\lambda^a$'s are $3\times 3$ Gell-Mann matrices and
$f_\pi\simeq 93$ MeV. Note that our result [eq. (\ref{effL0})] is
not contrary to that in Ref. \cite{BP93} since it does not give
any contribution to the imaginary part of the weak coupling
combination $N_{14}-N_{15}-2 N_{18}$. However, it is easy to see
that the effective lagrangian [eq. (\ref{effL0})] will give new
contributions to decays $K^+\to\pi^+\gamma\gamma$ and
$K^-\to\pi^-\gamma\gamma$, respectively, one can therefore expect
that the new structures in eqs. (\ref{effL1}) and (\ref{effL2})
with the complex Wilson coefficients $C^\pm_\gamma$ will induce a
possible charge asymmetry in $K^\pm\to\pi^\pm \gamma\gamma$
decays. Consequently, the interference between the amplitude for
$K^\pm\to\pi^\pm\gamma\gamma$ from eq. (\ref{effL0}) and the
absorptive part in eq. (\ref{amp1}) will give \beqn\label{asyemo}
{\delta\Gamma}^{\rm ~EMO}
&=&|\Gamma(K^+\to\pi^+\gamma\gamma)-\Gamma(K^-\to\pi^-\gamma\gamma)|^{\rm ~EMO}\nonumber\\
&=&\Frac{\alpha_{\rm EM}^2|G_8| m_{K}^5}{3\cdot 2^{8}\pi^5 f_\pi^2
}\left|{\rm
Im}C_\gamma^+\left(\frac{2}{3}a_1-a_2\right)\right|\nonumber\\
&&\times
\int^{(1-r_\pi)^2}_{4r_\pi^2}dz~\lambda^{1/2}(1,z,r_\pi^2)(r_\pi^2-1-z)z
~{\rm Im}F(z/r_\pi^2). \eeqn The first observation of
$K^+\to\pi^+\gamma\gamma$ has been reported by BNL E787
Collaboration \cite{E787}, and the branching ratio of the decay
has been measured \cite{E787, PDG02} \beq\label{Br}{\rm
Br}~(K^+\to\pi^+\gamma\gamma)=(1.1\pm0.3\pm 0.1)\times
10^{-6}.\eeq Thus we have \beq\label{delta1}
\left({\delta\Gamma\over 2\Gamma}\right)^{\rm
EMO}=(2.4\pm0.7)\times 10^6 \left|{\rm Im }C^+_\gamma
\left(\Frac{2}{3}a_1-a_2\right)\right|.\eeq

Our next task is to evaluate the magnitude of ${\rm Im}
C_\gamma^+$ and $(2/3 a_1-a_2)$ to check whether it is possible to
get a significant charge asymmetry of
$K^\pm\to\pi^\pm\gamma\gamma$ from eq. (\ref{delta1}). However,
since unknown constants $a_1$ and $a_2$ in eqs. (\ref{effL1}) and
(\ref{effL2}) are related to the low energy chiral dynamics, at
the present we have no model-independent way to give a reliable
determination of them. In the following we will estimate them
using naive dimensional analysis, within the chiral quark model,
and employing lattice calculation, respectively.

{\it Naive dimensional analysis}~~ As the order-of-magnitude
estimate, using naive dimensional analysis \cite{NDA}, we can
obtain \beq\label{NDAest}a_1\sim a_2\sim
f_\pi\frac{f_\pi}{\Lambda_{\chi}}, \eeq with $\Lambda_\chi=4\pi
f_\pi$ as the chiral symmetry spontaneously breaking scale. So we
get \beq\left|\left(\Frac{2}{3}a_1-a_2\right)\right|\sim a_1\sim
a_2\sim \Frac{f_\pi}{4\pi},\eeq and \beqn\label{NDAasy}
\left(\Frac{\delta \Gamma}{2\Gamma}\right)^{\rm EMO}\sim 1.8\times
10^4~{\rm GeV}~|{\rm Im}C_\gamma^+|. \eeqn

{\it The chiral quark model}~~ The chiral quark model has been
extensively used to study low energy hadronic physics involving
strong and weak interactions \cite{ERT90, PR91, BEF94, MKG99}. In
order to study the direct CP violation in decays $K\to\pi
\ell^+\ell^-$ ($\ell=e,~\mu$), this model has been employed by the
authors of Ref. \cite{DG02} to evaluate the bosonization of the
EMO, which corresponds to the $a_2$ part in eq. (\ref{effL0}). In
the same way as that in Ref. \cite{DG02}, one can obtain the
effective lagrangian corresponding to the $a_1$ part in eq.
(\ref{effL0}). This leads to \beq\label{CQMest}
a_1=\frac{f_\pi^2}{4 M_Q},\;\;\;\; a_2=\frac{3 M_Q }{8\pi^2},\eeq
where the constituent quark mass $M_Q$ could be set about 0.3 GeV
\cite{DG02}. Thus we get \beqn\label{CQMasy} \left(\Frac{\delta
\Gamma}{2\Gamma}\right)^{\rm EMO}= (1.6\pm0.4)\times 10^4~{\rm
GeV}~|{\rm Im}C_\gamma^+|. \eeqn

{\it Lattice calculation}~~ So far there is no direct lattice
calculation on $a_1$ and $a_2$. However, note that the $a_2$ part
in eq. (\ref{effL0}) will also contribute to transition $K\to \pi
\gamma^*\to\pi \ell^+\ell^-$, and the first lattice calculation of
the matrix element of the EMO, $\langle
\pi^0|Q^+_\gamma|K^0\rangle$, has been done in Ref. \cite{BLMM00}.
One therefore could determine the value of $a_2$ by comparing the
result from eq. (\ref{effL0}) and that from the lattice
calculation. In general, the matrix element of the EMO can be
parametrized in terms of a suitable parameter $B_T$ \cite{BCIRS99,
BLMM00}
\beq\label{emome1}\langle\pi^0|Q^+_\gamma|K^0\rangle=i\frac{\sqrt{2}e
Q_d }{16\pi^2m_K} B_T ~p^\mu_\pi p^\nu_K F_{\mu\nu}.\eeq On the
other hand, using the effective lagrangian in eq. (\ref{effL0}),
one can get
\beq\label{emome2}\langle\pi^0|Q^+_\gamma|K^0\rangle=i\frac{\sqrt{2}e
Q_d }{16\pi^2m_K} \frac{2 a_2 m_K}{f_\pi^2} ~p^\mu_\pi p^\nu_K
F_{\mu\nu} .\eeq  Since $B_T=1.18\pm0.09$ has been found in the
lattice calculation \cite{BLMM00}, from eq. (\ref{emome1}) and
(\ref{emome2}), we have \beq\label{a2}a_2=0.010\pm0.001~ {\rm
GeV}.\eeq Unfortunately, now we are not able to use the similar
way to extract any information on $a_1$ from the lattice
calculation. However, by comparing the value of $a_2$ in eq.
(\ref{a2}) with those from naive dimensional analysis [in eq.
(\ref{NDAest})] and the chiral quark model [in eq.
(\ref{CQMest})], one can find that they are of the same order of
magnitude. Meanwhile, $a_1$ from eqs. (\ref{NDAest}) and
(\ref{CQMest}) are also of the same order of magnitude. This leads
to the same order of magnitude estimates for the charge asymmetry
in eqs. (\ref{NDAasy}) and (\ref{CQMasy}), which are from naive
dimensional analysis and the chiral quark model, respectively.
Therefore, in general, \beq\label{asy2}\left(\Frac{\delta
\Gamma}{2\Gamma}\right)^{\rm EMO}\sim 1.0\times 10^4~{\rm
GeV}~|{\rm Im}C_\gamma^+|\eeq could be expected except in a
fine-tuning case in which there is an accidental cancellation
between $2/3a_1$ and $a_2$ (since we cannot reliably fix the
relative sign of $a_1$ and $a_2$ in a model-independent way).

In order to go further into the charge asymmetry in eq.
(\ref{asy2}), now one has to compute the imaginary parts of the
Wilson coefficients $C_\gamma^\pm$, which are related to the short
distance physics. In the SM it is easy to get ${\rm Im}C_\gamma^+$
from eqs. (\ref{HMO}) and (\ref{SMHMO}) as
\beq\label{cgammaSM}|{\rm Im}C_\gamma^+ |^{\rm SM}=\frac{3
G_F}{\sqrt{2}}(m_s+m_d)|{\rm Im}\lambda_t~C_{12}|,\eeq where
$\lambda_t=V_{td}V_{ts}^*.$  Due to the smallness of ${\rm
Im}\lambda_t\sim 10^{-4}$, this contribution from the SM to the
charge asymmetry in $K^\pm\to\pi^\pm \gamma\gamma$ is strongly
suppressed and could be negligible. Therefore in the following we
will turn our attention to physics beyond the SM.

Among the possible new physics scenarios, low energy supersymmetry
(SUSY) \cite{SUSY}, represents one of the most interesting and
consistent extensions of the SM. In generic supersymmetric models,
the large number of new particles carrying flavor quantum numbers
would naturally lead to large effects in CP violation and
flavor-changing neutral current (FCNC) amplitudes \cite{FCNC}.
Particularly, one can generate the enhancement of
$C^\pm_{\gamma,~g}$ at one-loop, via intermediate squarks and
gluinos, which is due both to the strong coupling constant and to
the removal of chirality suppression present in the SM. Full
expressions for the Wilson coefficients generated by gluino
exchange at the SUSY scale can be found in Ref. \cite{GGMS96}. We
are interested here only in the contributions proportional to
$1/m_{\tilde{g}}$, which are given by \beqn C_{\gamma,~{\rm
SUSY}}^\pm(m_{\tilde{g}})=\Frac{\pi\alpha_s(m_{\tilde{g}})}{m_{\tilde{g}}}
\left[(\delta^{\rm D}_{\rm LR})_{21}\pm(\delta^{\rm D}_{\rm
LR})^*_{12}\right]
F_{\rm SUSY}(x_{gq})\label{CEMO},\\
C_{g,~{\rm
SUSY}}^\pm(m_{\tilde{g}})=\Frac{\pi\alpha_s(m_{\tilde{g}})}{m_{\tilde{g}}}
\left[(\delta^{\rm D}_{\rm LR})_{21}\pm(\delta^{\rm D}_{\rm
LR})^*_{12}\right] G_{\rm SUSY}(x_{gq})\label{CCMO}, \eeqn where
$(\delta^{\rm D}_{\rm LR})_{ij}=(M^2_{\rm D})_{i_{\rm L}j_{\rm
R}}/m^2_{\tilde{q}}$ denotes the off-diagonal entries of the
(down-type) squark mass matrix in the super-CKM basis,
$x_{gq}=m^2_{\tilde{g}}/m^2_{\tilde{q}}$ with $m_{\tilde{g}}$
being the average gluino mass and $m_{\tilde{q}}$ the average
squark mass. The explicit expressions of $F_{\rm SUSY}(x)$ and
$G_{\rm SUSY}(x)$ are given in Ref. \cite{BCIRS99}, but noting
that they do not depend strongly on $x$,
 it is sufficient,  for our purposes, to
approximate $F_{\rm SUSY}(x)\sim$ $F_{\rm SUSY}(1)=2/9$ and
$G_{\rm SUSY}(x)\sim$ $G_{\rm SUSY}(1)=-5/18$. In any case it will
be easy to extend the numerology once $x_{gq}$ is better known.
Also the determination of the Wilson coefficients in eqs.
(\ref{CEMO}) and (\ref{CCMO}) can be improved by the
renormalization group analysis \cite{BCIRS99, BLMM00}. Then by
taking $m_{\tilde{g}}=500$ GeV, $m_t=174$ GeV, $m_b=5$ GeV, and
$\mu=m_c=1.25$ GeV, we will have \beq\label{CEMOnum} \left|{\rm
Im} C_\gamma^+\right|^{\rm SUSY}=2.4\times 10^{-4}{\rm
GeV^{-1}}~\left|{\rm Im} [(\delta^{\rm D}_{\rm
LR})_{21}+(\delta^{\rm D}_{\rm LR})^*_{12}]\right|. \eeq Using the
experimental upper bound on the branching ratio of $K_L\to\pi^0
e^+e^-$ measured by KTeV Collaboration \cite{KTeV00}, the lattice
calculation \cite{BLMM00} has given \beq\label{deltaplus}
\left|{\rm Im} [(\delta^{\rm D}_{\rm LR})_{21}+(\delta^{\rm
D}_{\rm LR})^*_{12}]\right|<1.0\times 10^{-3}~~~~(95\% ~{\rm
C.L.}). \eeq Thus from eqs. (\ref{asy2}), (\ref{CEMOnum}), and
(\ref{deltaplus}), the charge asymmetry in
$K^\pm\to\pi^\pm\gamma\gamma$ induced by the EMO in the
supersymmetric extensions of the SM could be bound as
\beq\label{SUSYasy} \left(\Frac{\delta
\Gamma}{2\Gamma}\right)^{\rm EMO}_{\rm SUSY}< {\rm a~few}\times
10^{-3},\eeq which  is significantly larger than the charge
asymmetries given in the SM \cite{DI98, DEIN95, BP93}.

In conclusion, we have studied the CP-violating charge asymmetry
induced by the electromagnetic penguin operators  in
$K^\pm\to\pi^\pm \gamma\gamma$ transitions, and supersymmetric
extensions of the SM may enhance the Wilson coefficients of these
operators, which leads to interesting phenomenology in this study.
It is found that the constrain imposed by experiments
\cite{KTeV00} on the SUSY parameter $\left|{\rm Im} [(\delta^{\rm
D}_{\rm LR})_{21}+(\delta^{\rm D}_{\rm LR})^*_{12}]\right|$ allows
a significant upper bound on the charge asymmetry given in eq.
(\ref{SUSYasy}). Our analysis shows that the charge asymmetry in
$K^\pm\to\pi^\pm\gamma\gamma$ up to $10^{-3}$ would be a signal of
new physics, and thus high-precision measurements of this
CP-violating observable might probe interesting extensions of the
SM. \vspace{0.5cm}
\begin{center}{\bf ACKNOWLEDGMENTS}\end{center}
The author wishes to thank G. D'Ambrosio for very helpful
communications. This work was supported in part by the NSF of
China under Grant No. 10275059.

\end{document}